\documentclass[letterpaper, 10 pt, conference]{ieeeconf}  

\IEEEoverridecommandlockouts                              
                                                          
\usepackage{cite}
\usepackage{amsmath,amssymb,amsfonts}
\usepackage{graphicx}

\usepackage{array}
\usepackage{lipsum}  

\usepackage{textcomp}
\usepackage{xcolor}
\usepackage{algorithm}
\usepackage{algpseudocode}
\usepackage{subcaption}

\usepackage{epsfig} 
\usepackage{hyperref}
\hypersetup{
    colorlinks=true,
    linkcolor=teal,
    filecolor=cyan,      
    urlcolor=teal,
    }
    
\usepackage{url}

\usepackage{color}

\newcommand{\blue}[1]{#1}  

\begin{document}
\hypersetup{bookmarksdepth=-2}

\title{Training Human-Robot Teams by Improving Transparency Through a Virtual Spectator Interface

    \thanks{$^{1}$ H. Qiang, W. Jo, L Robert and Dawn M. Tilbury are with the Robotics Department, University of Michigan, Ann Arbor, MI 48105, United States
    {\tt\small \{hqiang, wonse, lprobert, tilbury\}@umich.edu}
    }%
    \thanks{$^{2}$ M. AbuHijleh, S. Dallas and W. Louie are with the Department of Electrical and Computer Engineering, Oakland University, Rochester, MI 48309, United States
    {\tt\small \{abuhijleh, srdallas, louie\}@oakland.edu}
    }%
    \thanks{$^{3}$  D. Tilbury is with the Department of Mechanical Engineering and Department of Electrical Engineering and Computer Science, University of Michigan, Ann Arbor, MI 48105, United States
    {\tt\small tilbury@umich.edu}
    }%
    \thanks{$^{4}$ L. Robert is with the School of Information, University of Michigan, Ann Arbor, MI 48105, United States
    {\tt\small lprobert@umich.edu}
    }%
    \thanks{$^{5}$ K. Riegner and J. Smereka are with U.S. Army DEVCOM Ground Vehicle Systems Center (GVSC), United States
    {\tt\small \{kayla.l.riegner.civ, jonathon.m.smereka.civ\}@army.mil}}%
    \thanks{DISTRIBUTION STATEMENT A. Approved for public release; distribution is unlimited. OPSEC\#9049.}
}

\author{Sean Dallas\textsuperscript{2}, Hongjiao Qiang\textsuperscript{1}, Motaz AbuHijleh\textsuperscript{2}, Wonse Jo\textsuperscript{1}, Kayla Riegner\textsuperscript{5}, \\ Jon Smereka \textsuperscript{5}, Lionel Robert\textsuperscript{1,4}, Wing-Yue Louie\textsuperscript{2}, and Dawn M. Tilbury\textsuperscript{1,3,*}
}

\maketitle

\begin{abstract} 
After-action reviews (AARs) are professional discussions that help operators and teams enhance their task performance by analyzing completed missions with peers and professionals. Previous studies comparing different formats of AARs have focused mainly on human teams. However, the inclusion of robotic teammates brings along new challenges in understanding teammate intent and communication. Traditional AAR between human teammates may not be satisfactory for human-robot teams. To address this limitation, we propose a new training review (TR) tool, called the Virtual Spectator Interface (VSI), to enhance human-robot team performance and situational awareness (SA) in a simulated search mission. The proposed VSI primarily utilizes visual feedback to review subjects' behavior. To examine the effectiveness of VSI, we took elements from AAR to conduct our own TR, and designed a 1 $\times$ 3 between-subjects experiment with experimental conditions: TR with (1) VSI, (2) screen recording, and (3) non-technology (only verbal descriptions). The results of our experiments demonstrated that the VSI did not result in significantly better team performance than other conditions. However, the TR with VSI led to more improvement in the subjects' SA over the other conditions.
\end{abstract}
\vspace{4pt}



\section{Introduction} \label{sec:intro}
A common method used to support task improvement and training in teams is the after-action review (AAR). AAR is a structured process that helps teams reflect on what happened during a task, why it happened, and how to improve future team performance. AAR benefits teams by enhancing situational awareness (SA) and understanding of mission tasks \cite{us_army_leaderss_1993}. Through AAR, team members can provide feedback on how communication styles affected the task, share knowledge and insights they gained, and build a consensus on what occurred during the task. Computer-aided AAR tools have emerged in recent decades as a means to improve AAR for human teaming scenarios \cite{clark_user_2004, stone_inexpensive_2016, thurston2011generic, allen1994after, bitoun2013innovative, Goldberg1999Training}. By collecting data during team missions, and presenting visualizations to the team in the AAR, computer-aided AAR tools can help human team members better understand the contributions and limitations of themselves and their teammates, and ultimately improve the overall effectiveness of teams in future tasks.
While AAR has proven beneficial for human teams, incorporating robotic teammates introduces new complexities. Robots face difficulties in conveying their actions and effectively communicating their performance in a way that humans can interpret \cite{natarajan_human-robot_2023}. While AAR tools have proven effective in enhancing human collaboration, they have not been investigated for training humans to work together with robots in teams.

Our research adapts the virtual spectator system Louie \textit{et al.} created \cite{louievirtual}, to develop a new computer-aided AAR tool for human-robot teams (HRT), called the virtual spectator interface (VSI). The adapted VSI provides increased transparency into the perception, actions, and decision-making of robotic teammates over traditional computer-aided AAR tools through visualizations of the autonomy levels of agents, hazards the agents detect, and the user's associated actions.

In order to evaluate the effectiveness of our VSI, we distilled elements from AAR into a shorter training review (TR) and conducted a between-subjects study focusing on HRT performance. We hypothesized that subjects who experienced the VSI would show greater performance improvements compared to those using traditional TR methods. Our findings revealed that all teams experienced performance improvement after TR, without statistically significant differences between the three conditions. However, our findings showed that the VSI led to the best improvement in those subjects who had poor initial SA. We concluded that the VSI would be beneficial for individuals who initially had lower SA in their HRT.

\section{Background} \label{sec:backgrond} 
\subsection{After-Action Reviews}
An after-action review (AAR) is a professional discussion of an event, focused on performance standards, that enables teammates to discover for themselves what was supposed to happen, what actually happened, why it happened, how to sustain strengths and improve weaknesses \cite{morrison1999foundations}. The purpose of an AAR is to analyze the operation process and provide valuable feedback essential to correct training deficiencies.

\subsection{Computer-Aided After-Action Reviews}

With the advent of higher fidelity simulations and their rising use as training mediums, multimedia-aided AARs have transformed into a powerful means of enhancing AAR for simulation-based training environments. The Dismounted Infantry Virtual After-Action Review System \cite{clark_user_2004} and the Generic Engine for After-Action Review Scenarios \cite{thurston2011generic} form the basis and inspiration for what many future simulations incorporate into their computer-aided AAR tools today, such as task-specific playback functionalities, viewing modes, telemetry-based statistics, and overlays to indicate entity alignment and objectives \cite{clark_user_2004, stone_inexpensive_2016, thurston2011generic, allen1994after, bitoun2013innovative, Goldberg1999Training}.

User studies utilizing computer-aided AAR tools have shown how effective they can be. Stone \textit{et al.} developed a mine countermeasures simulation program to enable trainers to convey their own task-specific knowledge during AAR. The simulation program included the ability to plan optimal routes through the scenarios, summary screens with detailed statistics such as dwell times per object, and traditional replay capabilities like those found in \cite{clark_user_2004} and \cite{thurston2011generic}. In general, the use of this AAR tool was found to be effective, with a marginal increase in trainees search path accuracy and a large increase in both object identification and reporting accuracy of those objects \cite{stone_inexpensive_2016}.

\subsection{After-Action Review in Human-Robot Teams}
The introduction of robots into human teams restricts AAR due to the limited ability to communicate with human teammates \cite{natarajan_human-robot_2023}, resulting in low transparency of their actions and intent \cite{chen_assessment_2020}. Brewer \textit{et al.} highlights the needs of AAR with HRT and proposed Global AAR Technology (GAART), a tool which allows soldiers to view what happened during a mission from a birds-eye view accompanied by related statistics and overlays indicating technology usage, paths planned, and vehicle maneuvers \cite{wright_visualizing_2021}. 

Despite these efforts, recent computer-aided AAR present high-level visualizations that may not capture the intricacies of robotic behaviors and decision-making within HRT. This high-level visualization limits the ability to understand and analyze the actions and intentions of robotic teammates. To bridge this gap, it is crucial to develop more refined visualization tools that can offer deeper insights into the behavior and decision-making of robotic teammates, thereby, improving AAR for HRT.

\subsection{Virtual Human-Robot Interaction Spectating Tool}
Louie \textit{et al.} developed a virtual spectator system (VSS) to spectate virtual experiments \cite{louievirtual}, where soldiers interact with technology and teammates in a video game environment. The VSS allows for the observation and analysis of soldier interactions, including several key features such as graphs, map layers, events, overlays, and replays. These features enable spectators to review missions from a bird's-eye view, access data on soldier telemetry, visualize soldier actions and perception, and annotate key events for post-experiment analysis. The capability to visualize the intent of soldiers within the 3D space of the scene is valuable, as it enhances understanding and learning among team members during AAR \cite{yoo2024study}. This concept can be effectively extended to robotic teammates, who cannot contribute to AAR, by providing similar insights into their actions and decision-making process.




In this research, we aim to support training human teammates to work with robots in HRT by combining the features from the latest computer-aided AAR tools and transparency in robot behaviors gained from HRT spectating tools, to develop a more effective AAR tool for HRT.
\\

\section{Design of Virtual Spectator Interface (VSI)}  

\begin{figure*}[t]
    \centering    
   \vspace{5pt}

    \begin{subfigure}{1\linewidth} 
        \centering
        \vspace{5pt}
        \begin{subfigure}{0.36\linewidth}
            \centering
            \includegraphics[width=1\linewidth]{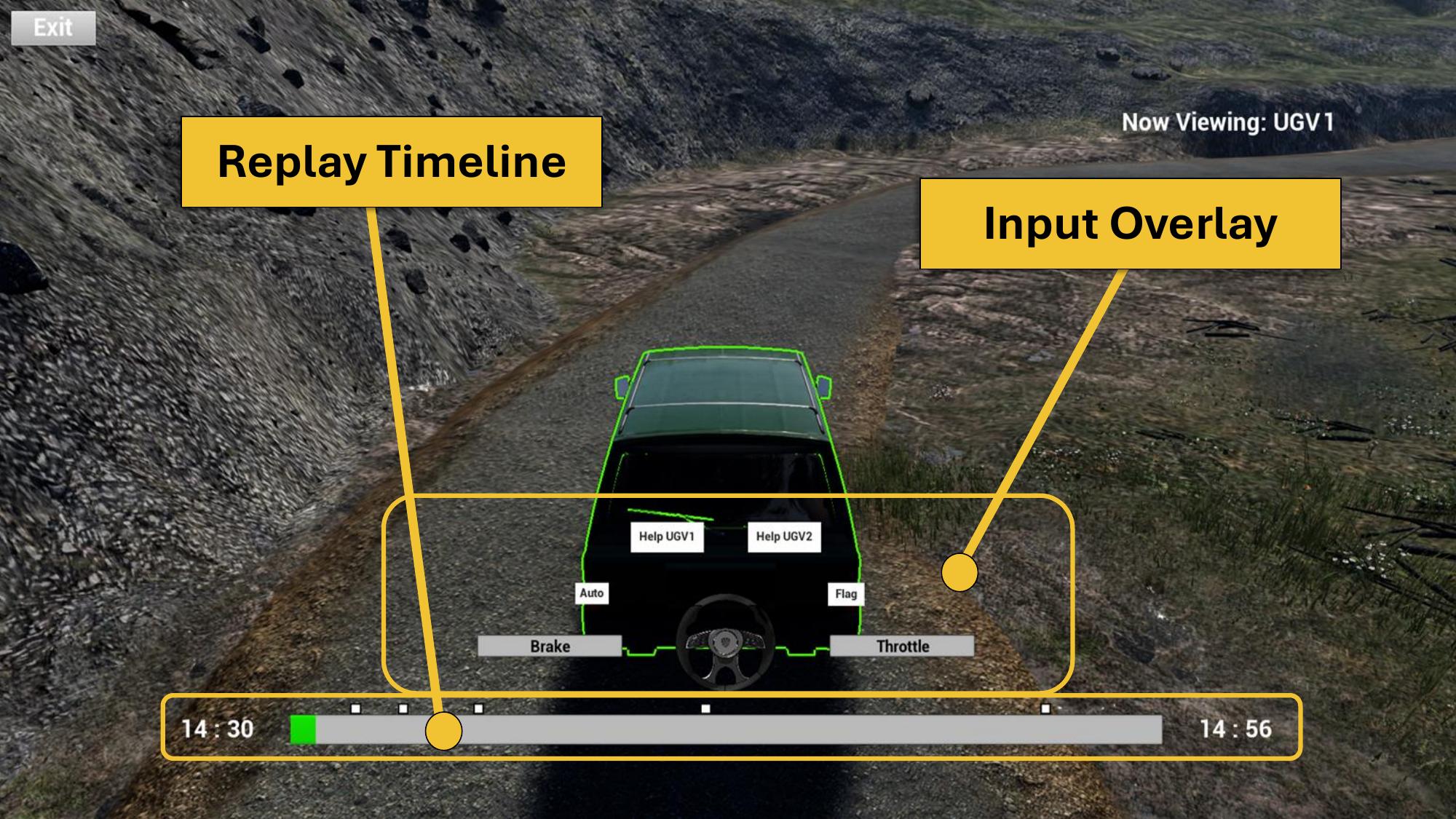}
            \caption{Input overlay and replay timeline}
            \label{fig:HUD Elements}
        \end{subfigure}        
        \begin{subfigure}{0.36\linewidth}
            \centering
            \includegraphics[width=1\linewidth]{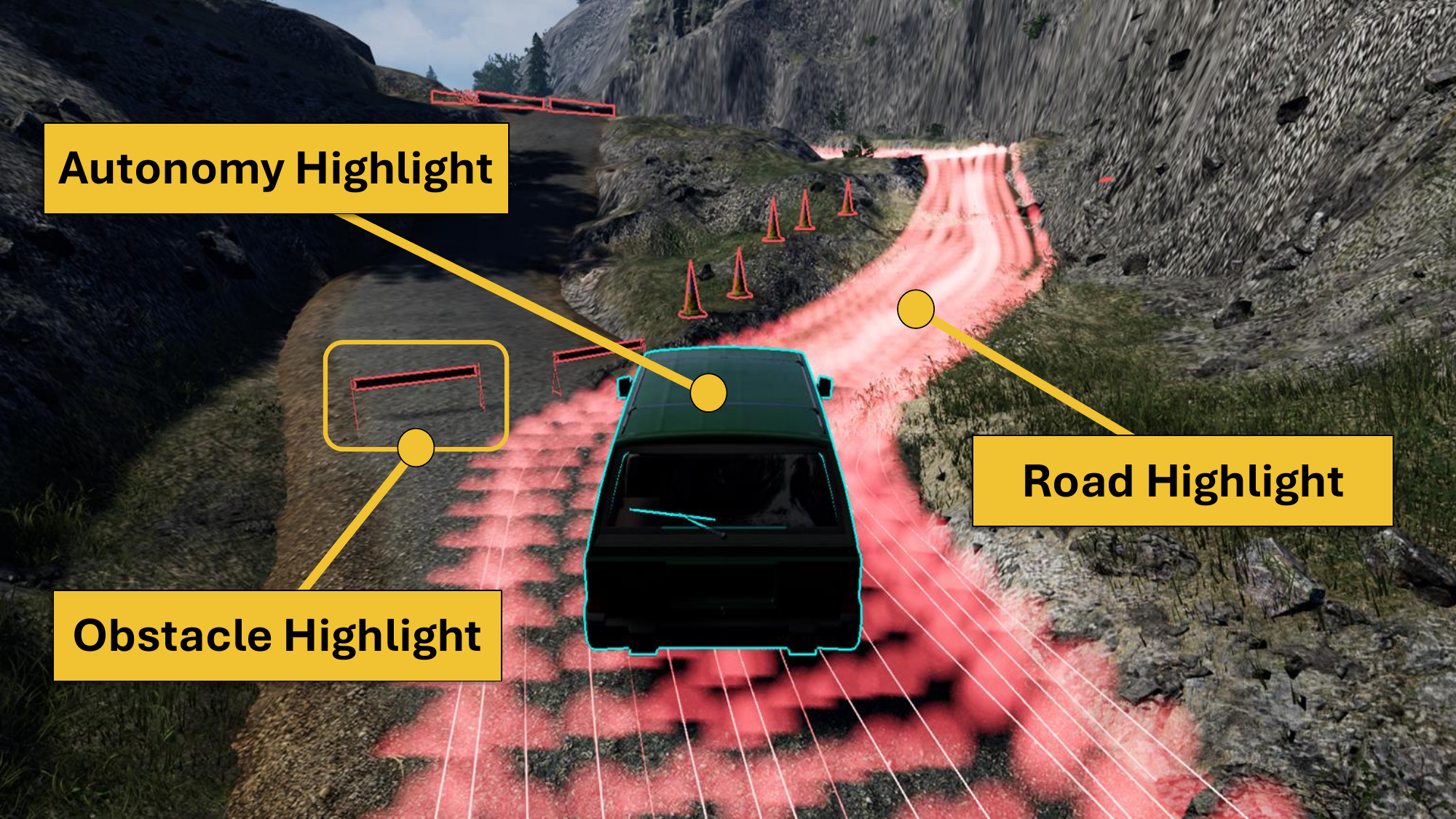}
            \caption{Highlighting obstacle, autonomy, and road}
            \label{fig:Overlay Elements}
        \end{subfigure} 
        \begin{subfigure}{0.26\linewidth}
            \centering     
            \begin{subfigure}{1\linewidth}
                \centering  
                \begin{subfigure}{0.485\linewidth}
                    \centering          
                    \includegraphics[width=1\linewidth]{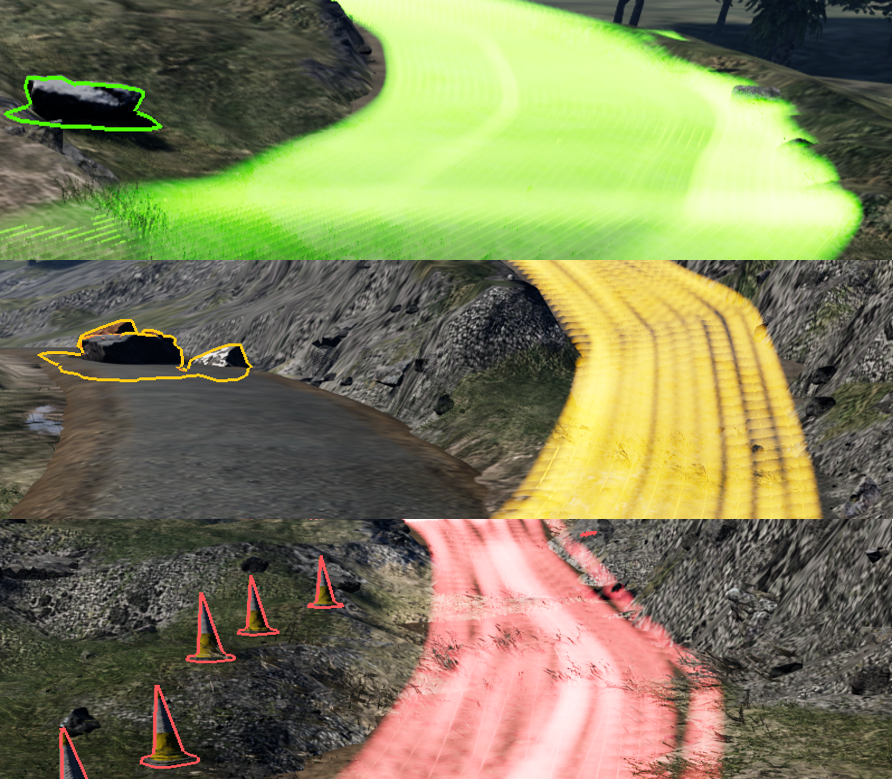}
                    \caption{HL on roads}
                    \label{fig:road_object_color_table}
                \end{subfigure} 
                \begin{subfigure}{0.485\linewidth}
                    \centering        
                    \includegraphics[width=1\linewidth]{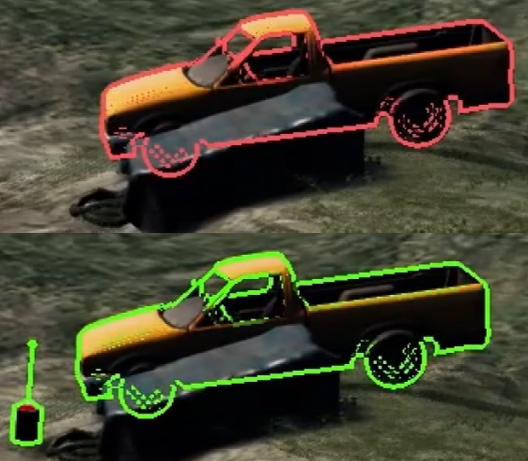}
                    \caption{OF status}
                    \label{fig:of_color_table}
                \end{subfigure} 
            \end{subfigure} 
            
            \begin{subfigure}{1\linewidth}
                \centering        
                \includegraphics[width=1\linewidth]{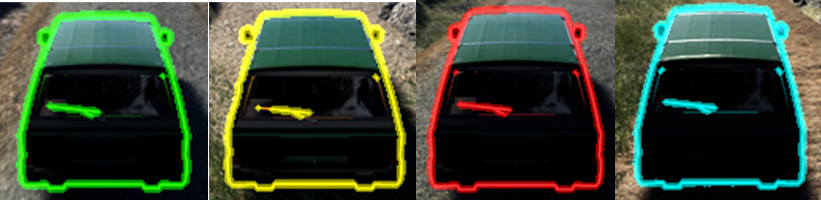}
                \caption{the UGV's autonomy }
                \label{fig:ugv_color_table}
            \end{subfigure}
            
        \end{subfigure}
    \end{subfigure}
    
    \caption{Examples of the Virtual Spectator Interface (VSI): (a) UI for input overlay and replay timeline and (b) UI for highlighting obstacle, autonomy, and road. The color used for highlighting represents, (c) the hazardous levels (HL) on roads, (d) the status of the flagging object, and (e) the status of UGV's autonomy. The supplementary video of the VSI can be found at \url{https://sites.google.com/umich.edu/mavric/projects/arc_sasi}.}
    \label{fig:vis_example}
        \vspace{-10pt}

\end{figure*}

The VSI is designed with a focus on transparency, clarity, and visibility, to help subjects comprehend the decisions of their robot teammates. Keeping in mind the need to minimize visual complexity \cite{carroll1987paradox, kostelnick2008visual}, elements were selected to enhance contrast, readability, and simplicity, with the aim to reduce cognitive workload. Additionally, the strategic use of color palettes further heightened visual clarity in order to facilitate rapid comprehension of presented interactions between the subject and the scenario \cite{treisman1980feature}. The VSI can be broken down into three main features: Input Overlay, Autonomy Outlines, and Object and Road Highlighting.

\subsection{Input Overlay} 
The input overlay in Fig.~\ref{fig:HUD Elements} aims to provide trainees with the opportunity to reflect on their inputs made during the previous mission. Reflection is a key component of what makes AARs successful \cite{keiser_meta-analysis_2021}, \blue{it enables} trainees to consolidate what they learned and deepen their understanding of the task. Within the input overlay we display the object flagging, braking, throttling, returning autonomy, taking-over autonomy, and steering wheel inputs. 


\subsection{Autonomy Outlines} 
The autonomy outlines feature was motivated by the necessity to enhance visual clarity and facilitate understanding the state of the UGVs autonomy \cite{rinehart2014sme,endsley2017here}. The status of object flagging targets and UGV's autonomy were also outlined during TRs according to the legend as shown in Fig.~\ref{fig:of_color_table} and \ref{fig:ugv_color_table}, respectively. 
The status of the UGV is indicated by different colors: green for full autonomy, orange for slowing down, red for stopped, and cyan for human takeover. In terms of object flagging, green means flagged and red means un-flagged.

This outline highlights these objects within the game-play footage, aiding users in quickly identifying and focusing on them \cite{treisman1980feature}. By providing clear visual indications of important elements, this outlining enhances operators' ability to analyze interactions and strategic decisions made during the mission.

\subsection{Road and Object Highlighting} The highlighting of roads and obstacles around hazards during replays was designed to support the task of determining where and when the UGV needed support. During the review, segments of the road and obstacles are highlighted with different colors according to the type of signs; obstacle signs are highlighted in green, caution signs in yellow, and stop signs in red (see Fig.~\ref{fig:Overlay Elements} and \ref{fig:road_object_color_table}).

This color-coded system provided operators with visual cues about the level of risk associated with different areas, enabling them to make informed adjustments to their understanding of the scenario during TR.

\section{Design of User Study} \label{sec:methodology} 
To investigate the effects of TR with our VSI on team performance and SA in a simulated HRT team conducting search missions, we conducted a 1$\times$3 between-subject experiment with three conditions: TR with the VSI ($C_{1}$), a screen recording ($C_{2}$), and a non-technology (verbal description) ($C_{3}$). In this section, we will explain the details of the mission and the design of the user study.

\subsection{Scenario and Task}
We utilized a simulated driving environment with Unreal Engine \cite{unrealengine}. The environment featured rough terrain for search missions where two UGVs navigated autonomously on different paths. The UGVs' cameras were streamed on a User Interface (UI) displayed on the large monitor as illustrated in Fig.~\ref{fig:hardware_system}. Subjects could see the status, communication messages, and video feeds from two UGVs on the left and right sides of the UI, respectively. 
During the experiment, subjects were asked to perform primary and secondary tasks. The primary task was to supervise the two UGVs and conduct two sub-tasks; Take Over (TO) sub-tasks and Object Flagging (OF) sub-tasks. The following are details of each sub-task of the primary task: 

\begin{figure}[t]
    \centering
    \includegraphics[width=0.90\linewidth]{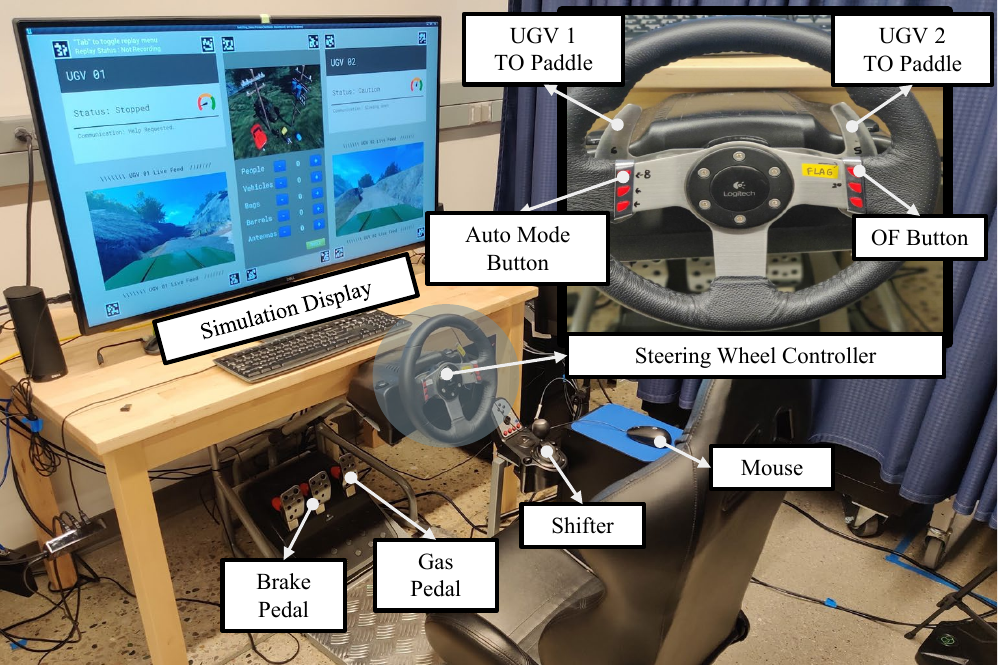}
    \caption{Details of a driving simulator controlling UGVs and the main experiment UI streaming the cameras of two UGVs and SR task on the large monitor.} 
    \label{fig:hardware_system}
    \vspace{-15pt}
\end{figure}

\begin{figure*}[t] 
  \centering
  \vspace{5pt}
  \includegraphics[width=0.95\textwidth]{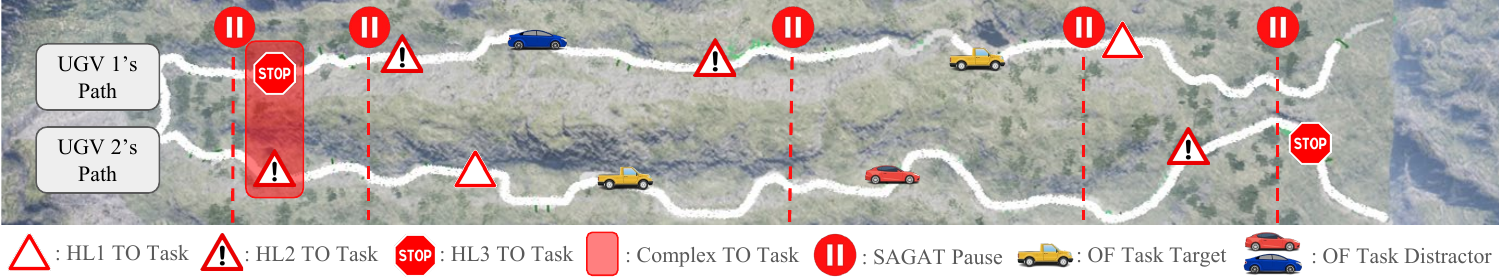}
  \caption{\blue{Map illustrating the distribution of takeover tasks, SAGAT pauses, and object flagging tasks }in Trial 2 map. More details of the maps used in Trial 1 and 2 can be found in this supplementary \blue{website}:   \url{https://sites.google.com/umich.edu/mavric/projects/arc_sasi}.}
  \label{fig:maps}
  \vspace{-15pt}
\end{figure*}

\subsubsection{Take Over (TO) Sub-task}
During navigation, UGVs encountered obstacles at three different hazard levels (HLs), which determined UGV behavior. The details of HL, UGV's behavior, and the subject's tasks are as follows:

\begin{itemize}
    \item \textbf{HL1}: This included ``Obstacle'' signs (e.g., \textit{``Car accident ahead''}, \textit{``Rock on the road''}, \textit{``Water on the road''}, and \textit{``Detour''}). The UGV displayed ``OK'' status and continued navigating autonomously at their original speed. No human intervention was necessary.

    \item \textbf{HL2}: This included ``Caution'' signs. The UGV displayed ``Caution'' status and continued navigating autonomously at a slower speed. The human subject could take over to drive faster or leave it autonomous.

    \item \textbf{HL3}: This included ``Stop'' signs. The UGV displayed ``Stopped'' status, stopped and communicated through text and audio ``Help requested''. The human subject must help by taking over the UGV.
\end{itemize}

During the TO sub-tasks, the video feed of the UGV filled the screen, and subjects manually controlled the UGV through the driving simulator. After passing obstacles, \blue{subjects had the option to return autonomy to the vehicle.}

\subsubsection{Object Flagging (OF) Sub-task}
The object flagging (OF) task required identifying yellow vehicles placed along the paths of the UGVs. Blue and red vehicles served as potential distractions and could be un-flagged if needed. When detecting an object, the UGV communicated ``Object Detected,'' paused for five seconds to process, and then decided whether to flag it. If flagged, a red and white pole appeared near the object. \blue{The UGVs had limited ability to accurately identify target objects. Therefore, subjects were required to carefully check whether an object was the target and unflag it if it had been incorrectly flagged.} The subjects could flag or un-flag an object using a button on the steering wheel.

\subsubsection{Secondary Task -- Spot Report (SR)}
Subjects also performed a secondary task known as the Spot Report (SR). The goal of this task was to create difficulty and distract the subjects' attention from the primary tasks \cite{10.1145/3610978.3640718}. In the Spot Report task, subjects viewed static images containing various categories of objects and were instructed to count the number of objects in each category (such as people, vehicles, bags, barrels, and antennas).

\subsection{Subjects Recruitment} 
We recruited 66 subjects (16 females, 49 males, and 1 non-binary; mean = 23.8 years old, SD = 6.9) at the University of Michigan and Oakland University to conduct the user experiments. All subjects had average knowledge of how to drive vehicles and did not have any visual or auditory disorders. Each subject was compensated \$20 for participating in the study. This study was approved by both University’s Institutional Review Boards (IRB); University of Michigan's IRB\#HUM00243123 and Oakland University's IRB\#IRB-FY2024-234. 

\subsection{Experiment Procedure}
We first explained the overall procedure and then obtained informed consent and demographic information. The subjects were then asked to watch task orientation videos and conduct a training session, to familiarize themselves with the sub-tasks (e.g., TO, OF, and SR tasks), and operate the driving simulator. Before starting a first trial, the experimenter started a screen recording program to record the subject's performance.

\subsubsection{Trial 1 ($T_{1}$)}
In $T_{1}$, there are five TO sub-tasks set on UGV 1's path and three TO sub-tasks set on UGV 2's path. One HL3 TO task on UGV 1's path and one HL2 TO task on UGV 2's path are intentionally set to happen at the same time for all subjects, called the ``Complex TO task''. There were two OF sub-tasks set on each UGV's path (see Fig.~\ref{fig:maps}). For the rest of the time, the UGVs drove autonomously and subjects were supposed to work on the SR task.

Throughout $T_{1}$, the simulation paused five times at certain points, and subjects were asked to fill out situation awareness global assessment technique (SAGAT) surveys \cite{endsley1988situation} for reporting their SA on each UGV. SAGAT was used to measure a subject's SA by asking three questions about their perception, comprehension of the current situation, and their projections of the future.

\subsubsection{Training Review (TR)} \label{sec:traing_review}
Following $T_{1}$, the experimenter performed the TR with a randomly selected experiment condition ($C_1$, $C_2$, or $C_3$). During the TR, the subjects reviewed four tasks from their previous $T_{1}$: ``Complex TO task'', ``HL1 TO task on UGV 2's path'', ``Second OF task on UGV 1's path'', and ``First OF task on UGV 2's path''. These tasks required human assistance to solve the problems on TO and OF sub-tasks, such as taking over UGV as soon as it requested help, and not taking any action for non-target objects encountered during the tasks. The details of the TR process based on experiment conditions are as follows:



\paragraph{For $C_1$}The experimenter first introduced the features of the VSI to subjects. Subjects then reviewed the four tasks in $T_{1}$ with VSI. Subjects were shown the road, obstacles, and UGV's autonomy highlighted with colors, and the control records subjects applied to the steering wheel during $T_{1}$. While reviewing the four tasks the experimenter verbally described what was supposed to be done for each example task.
\paragraph{For $C_2$} Subjects watched the screen-recorded video for TR. The experimenter showed the video clips of the four tasks to the subjects. At the same time, the experimenter verbally described what was supposed to be done in the four example tasks.
\paragraph{For $C_3$} Subjects had a non-technology TR. The experimenter verbally described what was supposed to be done in the four tasks. Then the experimenter described what the subjects actually did without any TR tool.

\subsubsection{Trial 2 ($T_{2}$)}
After the TR, the subjects were asked to perform the second trial ($T_{2}$) with a new map similar to $T_{1}$. In $T_{2}$, there are four TO sub-tasks, one ``Complex TO task'', and two OF sub-tasks on each UGV's path\blue{. The number of tasks was kept consistent between trials with the location and timing varied}. As with $T_{1}$, the simulation paused five times during $T_{2}$, where subjects were asked to answer SAGAT questions. For the rest of the time, subjects were supposed to work on SR tasks. Once both UGVs reached their destinations, the experiment ended.

\section{Measurements}  
\label{sec:measurement}
We evaluated the subject's task performance ($P$) and situational awareness ($SA$) by collecting data from the simulation and self-reporting questionnaires. The following sections will explain the measurement techniques that assess the effectiveness of VSI.

\subsection{Task Performance ($P$)} 
We calculated $P$ as: 

\begin{equation}\label{eq:performance}
    P = P_{TO} + P_{OF}
\end{equation}

    \noindent where $P_{TO}$ is the performance of the TO sub-task and $P_{OF}$ is the performance of the OF sub-task, which are combined due to them being the focus of the TR subjects' experience between trials. The details of $P_{TO}$ and $P_{OF}$ are measured as follows:

\subsubsection{Performance of TO task ($P_{TO}$)} 
We measured the total completion time by subtracting a standard completion time ($SCT$) from the subject's completion time in the trial to isolate the time spent on human intervention, providing insights into decision-making and efficiency. Thus, the total completion time represents the performance of the TO task ($P_{TO}$); faster/sooner is better performance. The $SCT$ in each trial was measured by running the UGVs through the fully autonomous navigation without stopping for any of the obstacles. $SCT_{T_{1}}$ is 783 seconds and $SCT_{T_{2}}$ is 552 seconds. 

\subsubsection{Performance of OF task ($P_{OF}$)} 
We calculated the total number of correct decisions made in the OF sub-tasks of each trial. If the subject flags the target object or un-flags the non-task object, they receive a point. If not, the subject loses a point. Thus, the total number of correct decisions represents the performance of the OF task ($P_{OF}$). The highest $P_{OF}$ in each trial is four points; higher is better performance. 

Then, the improvement in overall task performance ($\Delta P$) is calculated with the following equations:

\begin{equation}\label{eq:performance_improvement}
    {\Delta P} = (P_{T_{2}} - P_{T_{1}} ) / (P_{T_{1}}) \cdot 100\%
\end{equation}

\subsection{Situation Awareness ($SA$)}
We used modified SAGAT questionnaires to measure the subject's SA on each UGV \cite{endsley1988situation} five times in each trial. The SAGAT is composed of three questions for each UGV to measure (1) perception and (2) comprehension of the current situation, and (3) projection of future status. Each correct answer received 1 point, and the overall SA of each trial is calculated by summing up all the points \cite{SALMON2009490}, up to a maximum of 30 points for $SA$ at each trial.
Then, the improvement in SA ($\Delta SA$) is calculated with the following equations:

\begin{equation}\label{eq:sa_score_improvement}
    {\Delta SA} = (SA_{T_{2}} - SA_{T_{1}} ) / (SA_{T_{1}}) \cdot 100\%
\end{equation}

\section{Results and Analysis}  \label{sec:results}
We analyzed the objective and subjective responses of 66 subjects. All statistical analysis was performed using an open-source statistical package written in Python, Pingouin \cite{vallat2018pingouin}.

\subsection{Effects of Training Review (TR)} 
We conducted a paired sample t-test to investigate the effect of the TR on the subject's $P$ and $SA$ at $T_{1}$ and $T_{2}$ in all conditions ($C_{1}$, $C_{2}$, and $C_{3}$). 
From the results of the t-test, we found that the TR had a statistically significant impact on the improvement of the subject's $P$ ($T(130)=-6.41$, $p<.001$).
The mean and standard deviation (SD) of the subject's $P$ is $8.7\pm1.3$ in $T_{1}$ and $10.4\pm1.5$ in $T_{2}$.
Moreover, we found that the improvement in $SA$ was only borderline significant ($T(130)=-1.96$, $p=.05$) at $p=.05$.
The mean and SD of the subject's $SA$ is $22\pm6.1$ in $T_{1}$ and $24\pm5.2$ in $T_{2}$. Therefore, we can conclude that the TR has a statistically significant effect on improving the subject's task performance and effectively enhancing the understanding of the mission.

\subsection{Comparison of Experimental Conditions} 
For validating our hypotheses, we first performed a normality test to verify if the improvement percentage of the subject's $P$ and $SA$ between $T_{1}$ and $T_{2}$ followed a normal distribution. The results indicated that both the improvement percentage of the task performance and SA scores were normally distributed. Thus, we conducted a one-way Analysis of Variance (ANOVA) test with a significance level of $\alpha$ to $.05$ to explore the effects of each experimental condition ($C_{1}$, $C_{2}$, and $C_{3}$) on the subject's $\Delta P$ and $\Delta SA$.

\subsubsection{Improvement of Task Performance}
The results of the one-way ANOVA test indicated that there is no significant difference in the subject's $\Delta P$ ($F(2,63)=1.01$, $p=.37$ $\eta_{p}^{2}=.03$) on each condition. 
The mean and SD of $\Delta P$ at $C_{1}$, $C_{2}$, and $C_{3}$, is $24\pm23$, $16\pm18$, and $20\pm48$, respectively.

\subsubsection{Improvement of Situation Awareness}
The results of the one-way ANOVA test indicated that there is no significant difference in the subject's $\Delta SA$ ($F(2,63)=.45$, $p=.64$ $\eta_{p}^{2}=0.01$) on each condition. 
The mean and SD of $\Delta SA$ at $C_{1}$, $C_{2}$, and $C_{3}$ is $19\pm47$, $9.7\pm32$, and $20\pm40$, respectively.

Based on the one-way ANOVA test results of $\Delta P$ and $\Delta SA$, we can conclude that the proposed VSI-based TR does not have a significant impact on improving the subject's task performance and SA. 

\subsection{SA Comparison with Specific Groups}

\begin{figure}[t]
    \centering
    \vspace{5pt}
    \includegraphics[width=0.90\linewidth]{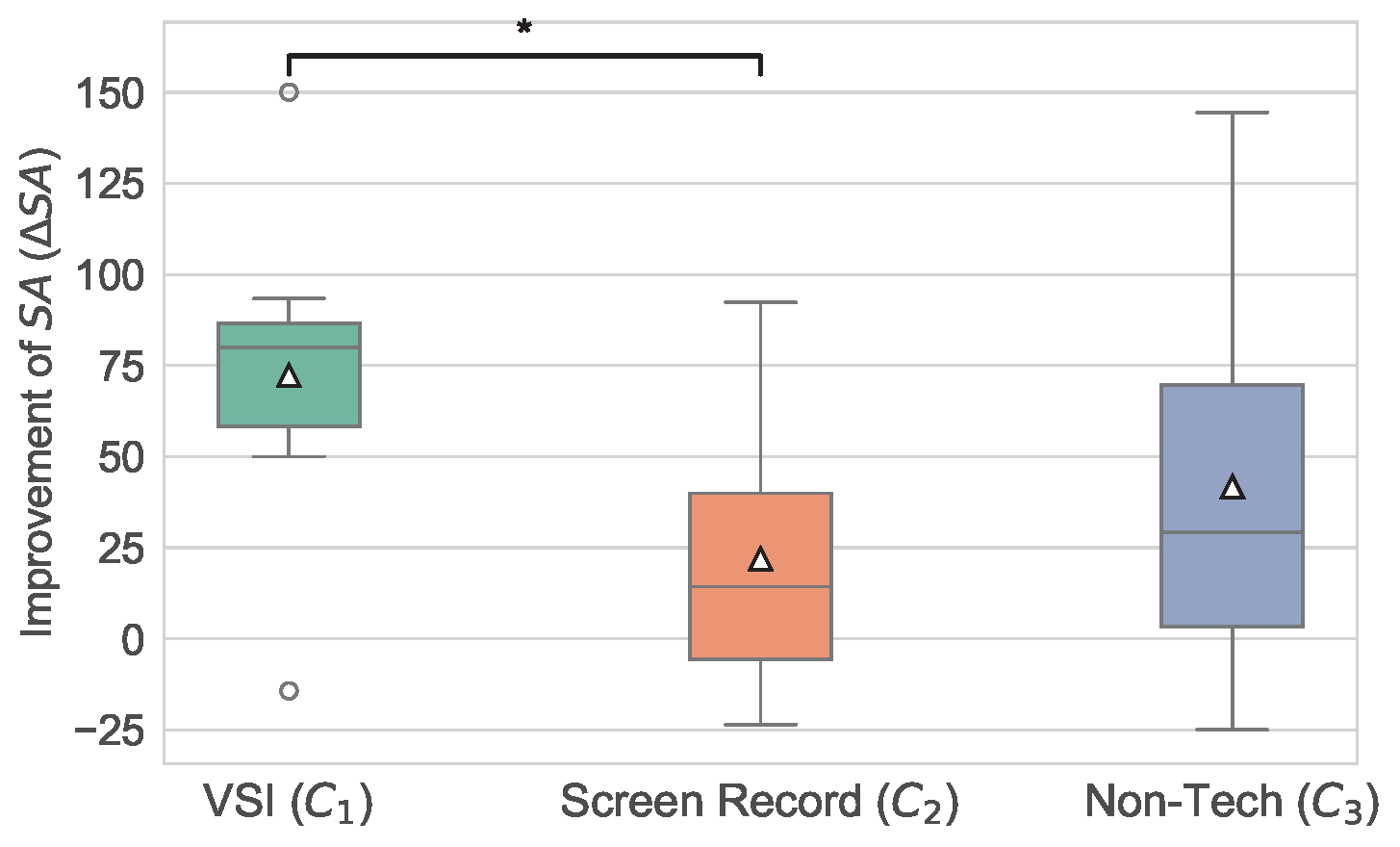}    
    \caption{The data distribution of $\Delta SA$ from a subgroup with scores lower than the mean of $SA$ in \blue{$T_{1}$} for all subjects. The white triangle indicates the mean.} 
    \vspace{-15pt}
    \label{fig:low_sa_group}
\end{figure} 

To thoroughly examine the impact of VSI on specific groups, we categorized subjects into two groups based on their $SA$ in $T_{1}$; 
(1) Group A ($n$=28): subject pool having lower $SA$ than the mean of $SA$ ($21.5$) in $T_{1}$ from all subjects.
and (2) Group B ($n$=32): subject pool having higher $SA$ than the mean of $SA$ in $T_{1}$ from all subjects. 
Fig.~\ref{fig:low_sa_group} illustrates the data distribution of the SA scores from Group A. 
We first conducted a Homoscedasticity test on each group to decide the statistical analysis method \cite{erceg2008modern}. For Group A, the Homoscedasticity result was $W=.30$, $p=.74$. For Group B, the result was $W=.26$, $p=.77$. The results indicate that the two groups do not have a Heteroskedasticity problem since neither $p$-value is significant, which means that both groups have similar variances.

Therefore, we conducted the one-way ANOVA test on each group and found \blue{a trend towards significance} between conditions ($F(2,25)=2.56$, $p=.10$, $\eta_{p}^{2}=.17$) from Group A, but there is no statistically significant difference between conditions from Group B ($F(2,35)=0.87$, $p=.43$, $\eta_{p}^{2}=.05$). 
We then conducted a post-hoc test using Turkey's method on Group A and found that there is a marginally significant difference in $\Delta SA$ between $C_{1}$ and $C_{2}$ ($p-tukey=.08$). The mean and SD of $\Delta SA$ in Group A is $72\pm49$, $22\pm39$, and $42\pm51$ at $C_{1}$, $C_{2}$, and $C_{3}$.

Therefore, we can conclude that although the proposed VSI does not directly impact all subjects' improvement in task performance and SA, it is more effective in enhancing the SA of individuals with lower SA during the mission than those with higher SA.

\section{Discussion} \label{sec:discussion}
This study investigated the impact of using our VSI on enhancing human performance in HRT through TR. We hypothesized that the VSI would lead to greater performance and SA improvements than traditional TRs. Our analysis demonstrated that all TR formats resulted in performance improvements ($T(130)=-6.41$, $p<.001$), which aligns with the literature as AARs increase performance in humans regardless of format \cite{keiser_meta-analysis_2021,villado2013comparative,tannenbaum_team_2013}. 

We also theorize that the input overlay of the VSI supported subjects' reflection on their actions during TR and led to consistent improvement in every condition across trials. However, only subjects who performed below the mean \blue{showed a trend towards improvement} in SA ($F(2,25)=2.56$, $p=.10$, $\eta_{p}^{2}=.17$). This selective improvement suggests that the VSI, which aimed to increase transparency in the robot's decision-making process and the subjects' behaviors during the review, may have particularly benefited those who initially had lower SA. 

In addition, we theorize that by providing clearer insights into the robot's operations through the autonomy, object, and road highlighting, the VSI may have helped these lower-performing subjects develop a better understanding and mental model of their robot teammates, which could contribute to their SA, as subjects understanding of the environment and where gaps in the UGVs autonomy exist would inform how and where they need to intervene \cite{bagnara_situation_2019}. 

Given the restricted ability of robotic teammates to communicate in teams and the resulting low transparency in action and intent \cite{natarajan_human-robot_2023, chen_assessment_2020} and the VSIs ability to improve SA after TR, designing an effective interface for future studies involving AARs in HRT should prioritize \blue{improving} transparency in robot behaviors to support learning. The interface should adapt the existing computer-aided AAR toolsets from tools like GAART \cite{wright_visualizing_2021} and include clear visualizations that depict the robot's decision-making process, including indicators of autonomy, object recognition, and any environmental variables that moderate the robot's ability to perform. By incorporating these elements, the interface can provide users with support to develop a comprehensive understanding of the robot's decision-making and operational context. 

Several limitations should be noted in the interpretation of these results. First, we did not include a control condition without TR, which could have helped determine whether the observed improvements were attributable to the TR process itself or merely due to increased familiarity with the simulation. Future studies should incorporate such a control condition to better isolate the effects of the TR. Additionally, the method of having each subject review their own footage introduced variability in the TRs. This discrepancy in review content may have contributed to the lack of statistically significant differences in performance and SA across the conditions. To address this, future research could standardize the review process or employ more consistent review criteria to ensure comparability across subjects.


\section{Conclusions and Future Works} \label{sec:conclusion}

In this study, we proposed a new training review (TR) tool called virtual spectator interface (VSI) and examined the effects of VSI on task performance and SA in human-robot team by comparing it with other TRs using traditional after-action review methods. Through user experiments, we found that TRs have positive impacts on subjects' performance and situation awareness (SA). However, when we compared the impact of different TR formats, we found there were no significant differences. We found a \blue{trend towards significance} between the TR with our VSI and the screen record condition when subjects had lower-than-mean situation awareness in the first trial. The result implies that the additional features of the VSI did not provide substantial benefits in this context. While advanced review mechanisms may hold potential, their effectiveness may vary and require further investigation.

Extending the duration of the simulated mission could provide insights into the long-term effects of TR with VSI. Future research should consider what information in a TR is most beneficial to improving SA and performance\blue{, and may also explore incorporating users' gaze patterns or physiological monitoring to provide deeper insights.} Conducting more extensive studies can help us better understand how involving more immersive technologies impacts the effectiveness of TR.

\vspace{-5pt}
\section*{Acknowledgement}
\blue{The authors wish to acknowledge the technical and financial support of the Automotive Research Center (ARC) and Immersive Simulation Directorate in accordance with Cooperative Agreement W56HZV-24-2-0001 U.S Army DEVCOM Ground Vehicle Systems Center (GVSC) Warren, MI.}

\bibliographystyle{IEEEtran}
\bibliography{references}






\end{document}